\documentclass[twocolumn,aps,prd,nofootinbib,amsmart]{revtex4-2}
 \usepackage{graphicx,amssymb}
 \usepackage[utf8]{inputenc} 

\usepackage{comment}
\usepackage[urlcolor=blue,colorlinks,breaklinks=true]{hyperref}

\PassOptionsToPackage{obeyspaces,spaces,hyphens}{url}

 \usepackage{amsmath}

\begin{document}
 	\def\half{{1\over2}}
 	\def\shalf{\textstyle{{1\over2}}}
 	
 	\newcommand\lsim{\mathrel{\rlap{\lower4pt\hbox{\hskip1pt$\sim$}}
 			\raise1pt\hbox{$<$}}}
 	\newcommand\gsim{\mathrel{\rlap{\lower4pt\hbox{\hskip1pt$\sim$}}
 			\raise1pt\hbox{$>$}}}

\newcommand{\be}{\begin{equation}}
\newcommand{\ee}{\end{equation}}
\newcommand{\bq}{\begin{eqnarray}}
\newcommand{\eq}{\end{eqnarray}}
 	
\title{Stochastic gravitational wave background generated by domain wall networks}

\author{D. Gr{\"u}ber}
\email[Electronic address: ]{david.grueber@astro.up.pt}
\affiliation{Departamento de F\'{\i}sica e Astronomia, Faculdade de Ci\^encias, Universidade do Porto, Rua do Campo Alegre 687, PT4169-007 Porto, Portugal}
\affiliation{Instituto de Astrof\'{\i}sica e Ci\^encias do Espa{\c c}o, Universidade do Porto, CAUP, Rua das Estrelas, PT4150-762 Porto, Portugal}

\author{L. Sousa}
\email[Electronic address: ]{lara.sousa@astro.up.pt}
\affiliation{Departamento de F\'{\i}sica e Astronomia, Faculdade de Ci\^encias, Universidade do Porto, Rua do Campo Alegre 687, PT4169-007 Porto, Portugal}
\affiliation{Instituto de Astrof\'{\i}sica e Ci\^encias do Espa{\c c}o, Universidade do Porto, CAUP, Rua das Estrelas, PT4150-762 Porto, Portugal}

\author{P.P. Avelino}
\email[Electronic address: ]{pedro.avelino@astro.up.pt}
\affiliation{Departamento de F\'{\i}sica e Astronomia, Faculdade de Ci\^encias, Universidade do Porto, Rua do Campo Alegre 687, PT4169-007 Porto, Portugal}
\affiliation{Instituto de Astrof\'{\i}sica e Ci\^encias do Espa{\c c}o, Universidade do Porto, CAUP, Rua das Estrelas, PT4150-762 Porto, Portugal}

\date{\today}

\date{\today}
\begin{abstract}
In this work we study the power spectrum of the Stochastic Gravitational Wave Background produced by standard and biased domain wall networks, using the Velocity-dependent One-Scale model to compute the cosmological evolution of their characteristic scale and root-mean-squared velocity. We consider a standard radiation + $\Lambda \rm CDM$ background and assume that a constant fraction of the energy of collapsing domain walls is emitted in the form of gravitational waves. We show that, in an expanding background, the total energy density in gravitational radiation decreases with cosmic time (after a short initial period of quick growth). We also propose a two parameter model for the scale-dependence of the frequency distribution of the gravitational waves emitted by collapsing domain walls. We determine the corresponding power spectrum of the Stochastic Gravitational Wave Background generated by domain walls, showing that it is a monotonic decreasing function of the frequency for frequencies larger than that of the peak generated by the walls that have decayed most recently. We also develop an analytical approximation to this spectrum, assuming perfect linear scaling during both the radiation and matter eras, in order to characterize the dependence of the amplitude, peak frequency and slope of the power spectrum on the model parameters. 
\end{abstract}

\maketitle
\section{Introduction} 	
The discovery of gravitational waves using the ground-based LIGO and Virgo observatories~\cite{LIGOScientific:2016sjg,LIGOScientific:2017vwq} has opened a new window into the cosmos. Gravitational radiation allows us to probe the dynamics of extreme astrophysical events, such as binary black hole mergers and neutron star collisions. In addition to these resolved sources, however, there is a stochastic gravitational wave background that arises from the cumulative effect of numerous unresolved and independent sources distributed across the universe. Pulsar Timing Arrays, operating in the nanohertz frequency band, have recently announced its detection for the very first time~\cite{NANOGrav:2023gor,EPTA:2023fyk,Reardon:2023gzh} and upcoming observatories, such as the Einstein telescope and the space-borne LISA constellation of satellites, will enable us to probe this background at higher frequencies.

Various astrophysical and cosmological phenomena --- such as supermassive black holes, inflation, cosmic strings, phase transitions (see e.g \cite{LISA:2022yao,LISACosmologyWorkingGroup:2022jok}) --- are predicted to generate stochastic gravitational wave backgrounds. These signals carry information about the sources that originated them and may then enable us to unveil the fundamental processes that shaped the early universe. The observed background is likely to have been generated by several of these sources and understanding each of these contributions is therefore essential in order to use the observational data to its full potential as a probe of the underlying physics.  

In this paper, we focus on the stochastic gravitational wave background generated by cosmic domain walls --- which were proposed as a potential explanation for the detected signal~\cite{NANOGrav:2023hvm} --- and perform a detailed characterization of the expected spectrum. Domain walls may arise in phase transitions in the early universe as a consequence of the spontaneous breaking of a discrete symmetry. These topological defects correspond to two-dimensional surfaces separating regions in space that have different vacuum expectation values. The existence of cosmic domain walls is predicted in various theoretical frameworks, including some grand unified theories and other models incorporating scalar fields (see e.g.~\cite{vilenkinshellardCosmicStringsOther}). However, direct observational evidence for their existence remains elusive. In fact, standard domain wall networks have an energy density that grows over time with respect to the cosmic background during the radiation and matter eras, which means that they would eventually dominate the energy budget of the universe in the absence of dark energy. This led to strong constraints on the energy scale of the domain-wall-forming phase transition, as walls that form at a scale of $\eta \gtrsim 1\, \rm MeV$ would leave detectable imprints on the Cosmic Microwave Background~\cite{sousaCosmicMicrowaveBackground2015, Zeldovich:1974uw}. This stringent limit, known as the Zel'dovich bound, may however be easily evaded if walls are unstable and decay before the time of decoupling~\cite{Larsson:1996sp} and, in fact, these biased domain wall networks are predicted to form in many relevant scenarios~\cite{Holdom:1982ew,Gelmini:1988sf,Coulson:1995nv,Avelino:2008qy,Hiramatsu:2012sc,Krajewski:2016vbr,Chen:2020wvu,Dunsky:2021tih}. Here, we will develop an analytical framework to compute the gravitational wave background generated throughout the cosmological evolution of stable and biased domain wall networks, using some of the techniques developed in the context of (semi-) analytical studies~\cite{Sousa:2013aaa,sousaStochasticGravitationalWave2014,Sousa:2020sxs} of the stochastic gravitational wave background generated by cosmic strings.

This paper is structured as follows. In Sec. \ref{sec:theoFramework}, we revisit the velocity-dependent one-scale model for the evolution of a network of cosmic domain walls (Sec.~\ref{sec:VOS}) and use it to develop a framework to describe the gravitational wave emission generated during its evolution (Sec.~\ref{sec:basicassumptions}). In Sec. \ref{sec:resTotalEnergy}, this framework is applied to calculate the evolution of the total energy density in gravitational waves. Section \ref{sec:resFullSpectrum} is then dedicated to the computation of the full stochastic gravitational wave spectrum, including the full numerical calculation (\ref{sec:resNumerical}), an analytical approximation to our results (\ref{sec:ResAnalytic}) and an estimate of the contribution of the gravitational waves emitted during the sudden collapse of the network due to a bias in the scalar field potential (\ref{sec:resBias}). In Sec. \ref{sec:comparisonLiterature} we compare our findings to previous computations of the stochastic gravitational wave background generated by domain wall networks in the literature. We then conclude in Sec. \ref{sec:conclusions}.

Throughout this paper, we will work in natural units, where $c=\hbar=1$, with $c$ being the speed of light in vacuum and $\hbar$ being the reduced Planck constant. In these units Newton's gravitational constant is given by $G=6.70711\cdot 10^{-57} \, \text{eV}^{-2}$. Moreover, we will use the cosmological parameters measured by the Planck mission \cite{planckcollaborationPlanck2018Results2020}, where the values of the density parameters for radiation, matter and dark energy are respectively given by $\Omega_{\rm r} = 9.1476\cdot 10^{-5}$, $\Omega_{\rm m} = 0.308$, $\Omega_\Lambda=1-\Omega_{\rm r} - \Omega_{\rm m}$ and the Hubble constant is $H_0=2.13 \cdot h \cdot 10^{-33} \, \rm eV$, with $h=0.678$, at the present time.

\section{Theoretical framework}\label{sec:theoFramework}

Current studies~\cite{Hiramatsu:2013qaa,Kitajima:2015nla,Krajewski:2016vbr,Chen:2020wvu,Dunsky:2021tih,Ferreira:2022zzo,Badger:2022nwo} of the Stochastic Gravitational Wave Background (SGWB) generated by domain wall networks are based on a model inferred from the field theory simulations of~\cite{Hiramatsu:2013qaa}. Therein, the total energy density in Gravitational Waves (GWs) and their spectrum is measured in lattice simulations and, under certain assumptions (that we will later discuss in more detail) a model for the full spectrum is built. Lattice simulations, however, have limitations: their dynamical range and resolution is severely limited. Moreover, since the thickness of moving walls is Lorentz contracted, the resolution of the walls is also lost in the relativistic limit, which leads to an artificial decrease of their velocity through the emission of scalar radiation~\cite{Peyrard:1984kdi,Hindmarsh:2014rka}. This is precisely the limit in which significant GW emission is expected to occur, and so this loss of resolution suggests that the measurement of GWs in these simulations may produce inaccurate results.

Here, we will follow a different approach and develop a semi-analytical framework to study the SGWB produced by domain walls. Semi-analytical models have been highly successful in predicting the SGWB generated by cosmic strings~\cite{Kibble:1984hp,Caldwell:1991jj,DePies:2007bm,Sanidas:2012ee,sousaStochasticGravitationalWave2013,sousaStochasticGravitationalWave2014,sousaFullAnalyticalApproximation2020} and in reproducing, after calibration, the results of numerical simulations~\cite{Blanco-Pillado:2013qja,Auclair:2019wcv}. Moreover, these semi-analytical frameworks have the advantage of being very versatile, enabling for instance the study of non-standard cosmological and defect scenarios~\cite{Sousa:2016ggw,Cui:2017ufi,Guedes:2018afo}.

In this section, we start by reviewing the Velocity-dependent One-Scale (VOS) model for the cosmological evolution of domain wall networks, which will serve as the basis for the framework we shall develop later in the section. All the underlying assumptions will be discussed in detail. 

\subsection{The velocity-dependent one scale model for cosmic domain walls}\label{sec:VOS}
The VOS model was first introduced for cosmic strings~\cite{martinsQuantitativeStringEvolution1996}, and later extended to accommodate cosmic domain walls and other topological defects \cite{avelinoDomainWallNetwork2011, Sousa:2011ew,Sousa:2011iu}. It describes the time evolution of a cosmic defect network in a homogeneous and isotropic Friedmann-Lemaitre-Robertson-Walker universe in terms of two macroscopic quantities: its Root-Mean-Squared (RMS) velocity $\bar{v}$ and a characteristic length scale $L$, defined in terms of the proper domain wall energy per unit area $\sigma$ and their energy density $\rho$ as $L=\sigma/\rho$. The model treats domain walls as infinitely thin surfaces, whose dynamics is described by the $2$-dimensional Nambu-Goto (or Dirac) action. Assuming that the network itself is homogeneous and isotropic on cosmological scales, averaging the Nambu-Goto equations of motion over the whole network and assuming that $\left<v^4\right>=\bar{v}^4$ ~\footnote{See~\cite{Avelino:2020ubr} for a discussion of the potential impact of this assumption.}, one finds that the evolution of $\bar{v}$ and $L$ is of the following form:
\begin{align} 
    \frac{dL}{dt} & =  (1+3\bar{v}^2)HL + \tilde{c}\bar{v} \label{eq:VOSL}  \\
    \frac{d\bar{v}}{dt} & =  (1-\bar{v}^2)\left(\frac{k_w}{L}-3H\bar{v}\right) \label{eq:VOSv}  \, .
    \end{align}
Here $H=\dot{a}/a$ is the Hubble parameter and $a$ is the cosmological scale factor. A phenomenological term (that we will discuss in more detail later), parameterized by $\tilde{c}$ and describing the energy loss experienced by the network, was also introduced in Eq.~(\ref{eq:VOSL}). Note that here we have neglected the impact of the frictional force caused by the scattering of the particles of the surrounding plasma on the walls' dynamics. This friction may significantly damp wall motion and lead to a denser network (see e.g.~\cite{avelinoDomainWallNetwork2011}) in the early universe. Domain wall dynamics, however, is expected to become effectively frictionless as the universe expands and becomes less dense.

The VOS model for domain wall networks has two free parameters that may be calibrated with simulations: the curvature parameter $k_w$ and the energy loss parameter $\tilde c$. We will work with the unique calibration $\tilde{c}=0.5$ and $k_w=1.1$ as given in~\cite{leiteScalingPropertiesDomain2011}, which provides an overall fit for the radiation-, matter- and dark-energy-dominated eras. Note that, although numerical simulations~\cite{martinsExtendingVelocitydependentOnescale2016} and recent semi-analytical studies~\cite{avelinoParameterfreeVelocitydependentOnescale2020,avelinoComparingParametricNonparametric2020,avelinoAnalyticalScalingSolutions2022} indicate that the calibration of these parameters is not unique and should, in fact, depend on the expansion rate, this single calibration actually provides accurate results since we will only consider a radiation + $\Lambda$CDM background.

As the universe expands, a domain wall network continuously loses part of its energy. This happens because increasingly large domain walls will enter the horizon  (here, and throughout the paper, we shall refer to the standard cosmological model horizon --- always of the order of the Hubble radius --- simply as horizon). Once this happens, walls detach from the Hubble flow and decay quickly, converting  their energy into scalar and gravitational radiation. This continuous collapse of domains contributes to further increase the characteristic length of the network, an effect that is captured by the last term in Eq. \eqref{eq:VOSL}. Alternatively, we may express the rate of energy loss directly as:
\begin{equation}
    \left.\frac{d\rho}{dt}\right|_{\rm loss} = \tilde{c}\bar{v}\frac{\rho}{L} = \tilde{c}\bar{v} \frac{\sigma} {L^{2}} \, .
    \label{eq:VOSLoss}
\end{equation}

For a decelerating power-law expansion of the universe, with $a(t)\propto t^\lambda$ and $0<\lambda<1$, the VOS equations have an attractor solution usually referred to as the linear scaling regime. In this regime, $L$ grows linearly with cosmic time, such that $L=\xi_\lambda t$, and $\bar{v}=\bar{v}_\lambda$ remains constant, where
\begin{equation}
    \xi_\lambda = \sqrt{\frac{k_w(k_w+\tilde{c})}{3\lambda(1-\lambda)}}\,, \quad
    \bar{v}_\lambda = \sqrt{\frac{k_w(1-\lambda)}{3\lambda(k_w+\tilde{c})}} \label{eq:linearv} \, .
\end{equation}
For the remainder of this article, we will mostly focus on the cosmologically most relevant cases of radiation and matter domination (corresponding to a value of $\lambda$ of $1/2$ and $2/3$ respectively). We will use a subscript `r' or `m' to refer to the values of the corresponding variables in the radiation and matter eras respectively and we define $\mathcal{A}_\lambda \equiv \tilde{c}\bar{v}_\lambda / \xi_\lambda^2$. Furthermore,  we will write the time evolution of the scale factor as $a(t)/a_0=\mathcal{C}_\lambda t^\lambda$ where $\mathcal{C}_{\rm r}\equiv\mathcal{C}_{1/2}=[4H_0^2 \Omega_{\rm r}]^{{1}/{4}}$, $\mathcal{C}_{\rm m}\equiv \mathcal{C}_{2/3} =[(9/4)H_0^2 \Omega_{\rm m}]^{{1}/{3}}$ and $a_0$ is the value of the scale factor at present time. Note that during the transition between the radiation- and matter-dominated eras the network temporarily leaves the scaling regime. The matter era, however, does not last long enough for scaling to be fully re-established before the onset of dark energy. Nevertheless, our results will show that assuming a matter-era scaling regime until the present time provides an excellent approximation for our purposes.

In this discussion we have assumed that domain walls are stable and that the networks survive until the present day. Notice, however, that the ratio between the energy density of a scaling domain wall network and that of the background is a growing function of time both in the radiation- and matter-dominated eras. As a result, these walls would in general be expected to leave strong observational signatures in the large-scale structure of the cosmos. Still, the absence of these strong imprints can have multiple explanations. Domain walls may have never formed or could have formed before or during cosmic inflation, therefore being pushed out of the Hubble horizon. On the other hand, domain walls that are formed late in the cosmic history could be light enough to evade current observational limits on domain wall tension. However, yet another solution to the problem has been discussed extensively in literature: networks of heavy domain walls could have formed early in the cosmic history but decayed shortly after. Such a decay can be caused by a bias in the scalar field potential that gives rise to the domain walls, for example if one of the vacuum states is initially more likely to be populated~\cite{Larsson:1996sp} or if the energy densities of the minima of the potential are slightly different~\cite{Gelmini:1988sf}. In the latter case, a volume pressure may counteract the surface tension of the domain walls. As the average size of the domains grows during their evolution, this volume pressure will eventually dominate over the surface tension, causing the domains of true vacuum to rapidly expand and fill all regions in the universe. As was shown in literature~\cite{Gelmini:1988sf,Larsson:1996sp,Avelino:2008qy}, the rapid decay of the false vacuum happens once the characteristic length $L$ approaches a value of order $\epsilon / \sigma$, where $\epsilon$ is the (small) difference in potential energy of the minima. For earlier cosmological times, bias has a negligible impact on domain wall network dynamics and, thus, their evolution is well described by the VOS model up to the time of disappearance of the walls.  We shall discuss the potential signature of biased domain walls on the SGWB later in this work.

\subsection{Basic assumptions on the emission of gravitational waves}\label{sec:basicassumptions}

Domain wall networks have been extensively studied using field theory numerical simulations~\cite{Press:1989yh,Garagounis:2002kt,Oliveira:2004he,Avelino:2005pe,leiteScalingPropertiesDomain2011,martinsExtendingVelocitydependentOnescale2016,Correia:2017aqf} and these show that collisions and intersections between walls associated to different domains are quite rare. This means that such collisions cannot be expected to give rise to a significant contribution to the SGWB. In fact, as explained in the previous section, the evolution of the network is mainly determined by the collapse of increasingly large sub-horizon sized domains under the effect of their tension. As these walls collapse, they evaporate and no longer contribute to the energy budget of the network. In the last stages of collapse, however, walls accelerate to ultra-relativistic velocities and should then be expected to emit gravitational- as well as scalar radiation in the process. Here, we will take into account that \textit{the SGWB generated by domain walls is expected to be sourced by collapsing domains} rather than by the relatively slow-moving super-horizon sized walls that are still part of the network. Note that, although ultra-relativistic velocities may also sometimes be reached on highly curved regions on super-horizon walls --- which would also naturally lead to an energy loss ---, this process is akin to the collapse of sub-horizon domains and may thus be described similarly.

Since simulations show that the collapse of sub-horizon sized walls is a relatively quick process, occurring on timescales comparable to or shorter than one Hubble time, we will also assume that \textit{domain walls decay effectively immediately on a cosmological timescale} once they detach from the Hubble flow. In this case, by drawing an analogy to small cosmic string loops~\cite{sousaStochasticGravitationalWave2014} (which also evaporate almost instantaneously on cosmological timescales), we assume that the gravitational radiation is generated instantaneously by the collapsing walls. Assuming that a constant fraction $\mathcal{F}\le 1$ of the total energy lost by the network is converted into gravitational radiation, the rate of emission of GWs at a time $t$ measured by an observer at a time $t_{\rm f}$ is given by

\begin{equation}
    \frac{d\rho_{\rm gw}}{dt}=\mathcal{F}\left.\frac{d\rho}{dt}\right|_{\rm loss} \left(\frac{a(t)}{a_{\rm f}}\right)^4\, ,
    \label{eq:gw-density}
\end{equation}
where $a_{\rm f}=a(t_{\rm f})$ and we have included the dilution caused by the expansion of the background.

Although Eq.~(\ref{eq:gw-density}) is sufficient to determine the total energy density in GWs, to fully characterize the SGWB spectrum one also needs to know how this energy is distributed in frequency. Following~\cite{sousaStochasticGravitationalWave2014}, this may be achieved by resorting to a Probability Density Function (PDF) $p(f)$. The spectral energy density in GWs measured at a time $t_{\rm f}$ is then given by:

\begin{equation}
    \frac{d\rho_{\rm gw}}{dt df}=\frac{d\rho_{\rm gw}}{dt} p(f)\,.
    \label{eq:spectral-gw-density}
\end{equation}
The exact shape of $p(f)$ is unknown, but there are a few well-motivated assumptions that can be made about it. Since the emission of gravitational radiation can either be sourced by domains once they become relativistic in the final stages of collapse or by regions of high curvature on their surface, we can define a maximal radius of curvature $R_{\rm max}$ for GW emission. Here, we will assume that \textit{this maximum radius is proportional to the characteristic length of the network: $R_{\rm max}(t)=\alpha L(t)$, where $\alpha \lsim 1$ is a constant}. We will also assume that \textit{the minimum frequency of the GWs at the time of emission $f_{\rm min[e]}$ is equal to $f_{\rm min[e]}=2/R_{\rm max}$.} As we do not expect significant emission of gravitational radiation from wall segments with a curvature radius larger than $R_{\rm max}$, we will assume that there is a minimal frequency $f_{\rm min}$ associated with it below which the emission of GWs can be neglected. An observer at $t_{\rm f}$ then measures a minimal frequency of
\begin{equation}
 f_{\rm min}(t)=f_{\rm min[e]}(t)\frac{a(t)}{a_{\rm f}}=\frac{2}{R_{\rm max}(t)}\frac{a(t)}{a_{\rm f}}=\frac{2}{\alpha L(t)}\frac{a(t)}{a_{\rm f}} \,,
    \label{eq:fmin}
\end{equation}
where we have used the fact that $f(t_{\rm f})=(a(t)/a_{\rm f})f_{\rm e}$. 

The simplest PDF one may assume is a Dirac-$\delta$ function of the form:

\begin{equation}
    p(f)=\delta(f-f_{\rm min}) \,,
    \label{eq:pdfDelta}
\end{equation}
which corresponds to the unrealistic assumption that the gravitational radiation emitted by a collapsing domain wall has a frequency exactly given by $f_{\rm min}$. However, as a wall emits GWs, its energy should decrease until they eventually evaporate (although here we assume that this happens effectively immediately), so in general we should expect walls to emit GWs with frequencies $f\ge f_{\rm min}$.  By again drawing an analogy with small cosmic string loops, we would write~\cite{sousaStochasticGravitationalWave2014}:
\be
p(f)=p(R)\left|\frac{dR}{df_{\rm e}}\right| \frac{df_{\rm e}}{df} \theta(f-f_{\rm min})\,,
\ee
where $f_{\rm e}$ is the emitted frequency, $\theta$ is the heaviside step function, $R \propto \sqrt E$ where $E$ is the domain wall energy, and $p(R)=dE/dR \propto R$. Assuming that $f_{\rm e}=2/R$, its distribution should be such that $p(f) \propto f^{-3}  \theta(f-f_{\rm min})$. Note however that cosmic string loops survive for several oscillations while emitting GWs, which makes the distribution of energy in frequency during the evaporation process easier to predict. For domain walls this is not the case: due to their higher dimensionality, walls will in general self-intersect and evaporate upon first collapse without executing a quasi-periodical motion. This means that this direct generalization of the results of cosmic string loops would be questionable and that understanding what happens in these final stages of collapse (by means of numerical simulations) is crucial to determine the PDF. Given this, we will assume a more general power-law PDF that allows for the emission of frequencies larger than $f_{\rm min}$ with some arbitrary power $f^\nu$:
\begin{equation}
    p(f) = - \frac{\nu+1}{f_{\rm min}^{\nu+1}}f^\nu\theta(f-f_{\rm min}) \, . 
    \label{eq:PDFnu}
\end{equation}
Here we shall treat $\nu$ as a free parameter with the only limitation that $\nu < -1$, in order for the PDF to be normalizable. In fact, as one can easily verify, Eq. \eqref{eq:PDFnu} fulfills the condition $\int_0^\infty p(f) df=1$.

\section{Energy density in gravitational waves}\label{sec:resTotalEnergy}
In this section, before venturing into a full characterization of the SGWB spectrum generated by domain walls, we study how the total energy density of GWs is expected to evolve in a cosmological background. This study can be performed with minimal assumptions, just requiring that a constant fraction of the energy lost by the network goes into GWs (no assumptions regarding its distribution in frequency are necessary). The results presented here, then, are quite general and should therefore help clarifying the potential role of domain walls as contributors to the SGWB.

\begin{figure} 
	         \begin{minipage}{1.\linewidth}  
               \rotatebox{0}{\includegraphics[width=1\linewidth]{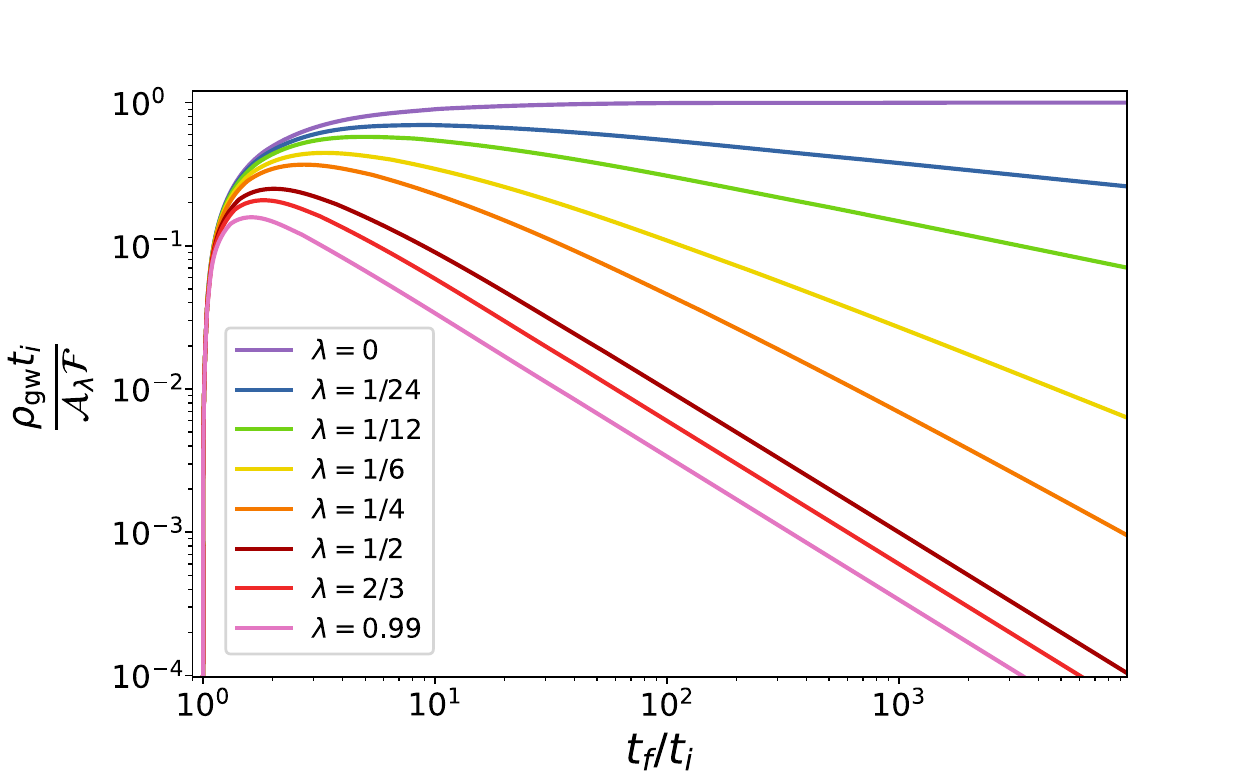}}
         \end{minipage}
	\caption{Evolution of the total energy density of gravitational waves generated by a domain wall network for different expansion rates $\lambda$. Here, we assume that $\mathcal{F}$ is a constant and a power-law expansion of the cosmological scale factor, such that $a\propto t^\lambda$.}
	\label{fig:totalrhogw}
\end{figure}

Assuming a power-law evolution of the scale factor and that the domain wall network is in the corresponding linear scaling regime, Eq. \eqref{eq:gw-density} may be rewritten as:
\begin{equation}
    \frac{d\rho_{\rm gw}}{dt} = \mathcal{A}_\lambda \mathcal{F} \sigma \frac{t^{4\lambda-2}}{t_{\rm f} ^{4\lambda}} \,.
\end{equation}
The total energy density in GWs measured by an observer at a time $t_{\rm f}$ is then given by:
\begin{equation}
    \rho_{\rm gw} (t_{\rm f})  = \int_{t_{\rm i}}^{t_{\rm f}} \frac{d\rho_{\rm gw}}{dt} dt 
     =  \sigma \frac{\mathcal{F}}{t_{\rm f}}\frac{\mathcal{A_\lambda}}{4\lambda-1}\left[1-\left(\frac{t_{\rm i}}{t_{\rm f}}\right)^{4\lambda-1}\right]\,
     \label{eq:totalrhogw}
\end{equation}
where we have assumed that the emission of GWs starts at a time $t=t_{\rm i}$. In Fig.~\ref{fig:totalrhogw}, the evolution of the energy density of gravitational radiation for different expansion rates is plotted assuming a constant $\mathcal{F}$. Therein, it is shown that for $\lambda>0$, after a short period of fast growth, the GW energy density decreases over time. In fact, by analysing Eq.~(\ref{eq:totalrhogw}), one can show that, asymptotically, $\rho_{\rm gw} \propto t_{\rm f}^{-1}$ for $\lambda>1/4$ (which includes the cosmologically relevant radiation and matter backgrounds) and $\rho_{\rm gw} \propto t_{\rm f}^{-4\lambda}$ for $\lambda<1/4$, while $\rho_{\rm gw} \propto \log(t_{\rm f})/t_{\rm f}$ for $\lambda=1/4$. Interestingly, although in a non-expanding background ($\lambda=0$) $\rho_{\rm gw}$ grows with time, one has that, for $t_{\rm f}\gg t_{\rm i}$, $\rho_{\rm gw}\to \mathcal{A}_0 \mathcal{F}/t_{\rm i}$ and remains roughly constant over time. This is explained by the fact that, in a linear scaling regime, the network becomes less dense over time and thus the energy lost by the network is smaller and smaller as time goes by. In an expanding background, the combination of this overall decrease in the rate of emission of gravitational waves with the dilution caused by expansion means that the total energy density of gravitational radiation should asymptotically decrease over time. As a matter of fact, as illustrated in  Fig.~\ref{fig:totalrhogw}, this decreasing regime is reached faster for larger $\lambda$ and, as a result, a smaller $\rho_{\rm gw}$ is measured at all times by observers in backgrounds with faster expansion rates.

Note that, although these results were derived assuming that a constant fraction of the energy lost by the network goes into GWs, its main conclusion --- namely that the total energy density of gravitational radiation generated by a domain wall network decreases over time in an expanding universe, after a relatively short-lived initial increase --- should hold quite generally. It is straightforward to show that having an increasing or constant GW energy density necessarily requires that the fraction of the energy lost by the network that is converted into gravitational radiation grows over time as $\mathcal{F}\propto t^s$, with $s\ge$ 1. However, one such scenario could only be transient as the energy density that goes into GWs naturally cannot exceed the energy lost by the network (notice that $\mathcal{F}\le 1$ should be verified at all times). On the other hand, assuming a decreasing $\mathcal{F}$ would necessarily lead to a faster decrease of $\rho_{\rm gw}$ over time than in the constant $\mathcal F$ case.

Let us now consider a biased domain wall network that decays at a time $t_\star$ for which $a_\star=a(t_\star)$. The total GW energy density generated by the scaling domain wall network (ignoring, for now, the emission of gravitational waves during the decay of the network itself), in units of the critical density  $\rho_{\rm crit} = 3H_0^2 / (8\pi G)$, measured by an observer at the present time is given by:

\be 
\left.\frac{\rho^{\rm r}_{\rm gw}}{\rho_{\rm crit}}\right|_{t=t_0}=\frac{16}{3}\pi \mathcal{F} G\sigma \frac{\Omega_{\rm r} ^{1/2}}{H_0}\mathcal{A}_{\rm r}\left(\frac{a_\star}{a_0}\right)^2\,,
\label{eq:totrhorad}
\ee 
for $a_\star<a_{\rm eq}$, where $a_{\rm eq}=a(t_{\rm eq})$ and $t_{\rm eq}$ is the time of radiation-matter equality and we have assumed that GW emission started at a time $t_{\rm i}\ll t_{\rm f}$, and

\bq 
\left.\frac{\rho^{\rm m}_{\rm gw}}{\rho_{\rm crit}}\right|_{t=t_0} & = & \frac{16}{3}\pi \mathcal{F} G\sigma \frac{\Omega_{\rm r} ^{5/2}}{H_0\Omega_{\rm m}^2}\mathcal{A}_{\rm r} + \label{eq:torrhomat}\\
&  \hspace{-1.2 cm}+ & \hspace{-0.8 cm}\frac{12}{5}\pi \mathcal{F} G\sigma \frac{\Omega_{\rm m}^{1/2}}{H_0} \mathcal{A}_{\rm m} \left(\frac{a_\star}{a_0}\right)^{5/2} \left[1-\left(\frac{a_{\rm eq}}{a_\star}\right)^{5/2}\right]\,,\nonumber
\eq
for $a_\star\ge a_{\rm eq}$. In both cases, one may see that the fractional contribution of the GWs emitted by domain walls to the energy density is a growing function of $t_\star$ (assuming that the other parameters are fixed). This is explained by the fact that the ratio between the energy density of domain walls and the background density rapidly grows with time both in the radiation and matter eras and thus $\rho_{\rm gw}$ is dominated by late-time GW emissions.

Let us consider stable walls that survive until the present day. The gravitational waves generated during the radiation era provide a contribution to the energy density of the universe at the present time of
\begin{equation}
    \left.\frac{\rho_{\rm gw}^{\rm r}}{{\rho_{\rm crit}}}\right|_{t=t_0} \frac{\sigma_{\rm Z}}{\mathcal{F} \sigma} = \mathcal{O}(10^{-14}) \,, 
\end{equation}
where we have introduced $\sigma_{\rm Z} = (10^{6} \, \text{eV})^3$ which corresponds to the mass of the heaviest walls that are still compatible with the Zel'dovich bound. On the other hand, considering gravitational wave emission during the matter dominated era, the fractional contribution is significantly larger:
\begin{equation}
    \left.\frac{\rho_{\rm gw}^{\rm m}}{{\rho_{\rm crit}}}\right|_{t=t_0} \frac{\sigma_{\rm Z}}{\mathcal{F} \sigma} =  \mathcal{O}(10^{-6}) \, . 
    \label{eq:densityMatter}
\end{equation}
This means that the GWs generated by stable domain walls compatible with the current CMB bounds~\cite{Sousa:2015cqa} provide, at most, a relative contribution of $10^{-6}$ to the total energy density of the universe, which indicates that, in general, we should not expect GW constraints on $\sigma$ to lead to significant improvements over the Zel'dovich bound. As a matter of fact, such a contribution would only arise in the unrealistic scenario in which all the energy lost by network goes into GWs (i.e. $\mathcal{F}$= 1) and, quite generally, this amplitude can be significantly suppressed for a smaller efficiency of GW emission $\mathcal{F} \ll 1$.

\section{Spectrum of the Stochastic Gravitational Wave Background}\label{sec:resFullSpectrum}
In this section, we will perform the full numerical characterization of the SGWB generated by domain wall networks in a $\rm \Lambda$CDM background and develop an accurate analytical approximation to describe these results. This SGWB may be characterized by the spectral density in GWs in units of $\rho_{\rm crit}$, given by
\be
\Omega_{\rm gw}(f) = \frac{1}{\rho_{\rm crit}}\frac{d\rho_{\rm gw}}{d\log f} = \frac{f}{\rho_{\rm crit}} \int_{t_{\rm i}}^{t_{\rm f}} \frac{d\rho_{\rm gw}}{dt df} dt  \,.
\label{eq:defOmega}
\ee
Here $t_{\rm f}={\rm min}(t_\star,t_0)$ and $t_{\rm i}$ denotes the time in which significant GW emission starts. Here, as is usually done for cosmic strings, we will assume that $t_{\rm i}$ corresponds to the time in which friction becomes negligible to the dynamics and not the time of creation of the domain wall network~\footnote{Note however that since the largest contribution to the SGWB comes from later cosmological times, the choice of $t_{\rm i}$ should not have a significant impact on observational constraints derived using our results.}. The SGWB generated by domain walls measured at the present time may then be found by setting $a_{\rm f}=a_0$ in Eq.~(\ref{eq:gw-density}) and integrating Eq. \eqref{eq:spectral-gw-density}.

.
\subsection{Full numerical computation in a $\rm \Lambda$CDM background}\label{sec:resNumerical}
To fully characterize the SGWB spectrum produced by domain wall networks in a $\rm \Lambda$CDM background, as a first step, we start by solving numerically the VOS Eqs.~(\ref{eq:VOSL}) and~(\ref{eq:VOSv}), coupled to the Friedmann equation, in order to get the realistic evolution of $L(t)$ and $\bar{v}(t)$. This enables us to compute the energy lost by domain wall networks throughout their evolution in a realistic cosmological background.

\begin{figure} 
	         \begin{minipage}{1.\linewidth}  
               \rotatebox{0}{\includegraphics[width=1\linewidth]{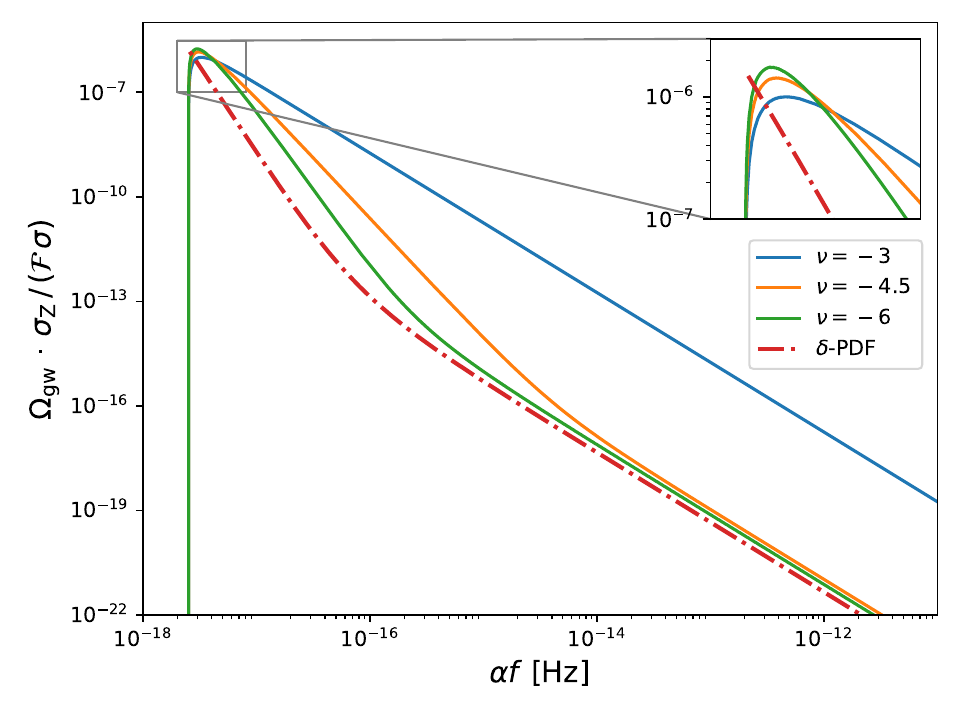}}
         \end{minipage}
	\caption{Stochastic Gravitational Wave Background generated by domain walls that survive until the present time. Here, $\Omega_{\rm gw}(f)$ is determined numerically as prescibed in Eq.~\eqref{eq:omegaNuNumerical} for different values of $\nu$ (solid lines). The spectrum resulting from the Dirac-$\delta$ PDF, as given in Eq. \eqref{eq:numericalDelta}, is also plotted (dash-dotted line). The vertical axis is re-scaled by a factor of $\sigma_{\rm Z}/(\mathcal{F} \sigma)$, so that the spectra correspond to domain walls with $\sigma = \sigma_{\rm Z}$ (saturating the Zel'dovich bound) and a maximum efficiency of gravitational wave emission of $\mathcal{F}=1$. The frequency on the horizontal axis is re-scaled by $\alpha$, since a lower value of $\alpha$ will merely shift the spectrum towards higher frequencies.}
	\label{fig:numericalNus}
\end{figure}

Let us first consider the case in which $p(f)$ is given by the $\delta$-function in Eq.~\eqref{eq:pdfDelta}. In this case, the integration in Eq.~(\ref{eq:defOmega}) yields a spectrum of the form:
\begin{equation}
    \Omega_{\rm gw}(f) = \frac{\pi\tilde{c} \mathcal{F}  G  \sigma }{6 H_0^2}     \frac{\bar{v}(t_{\rm min})L(t_{\rm min})^2 }{H(t_{\rm min})}(\alpha f)^4 \theta \left( f - f_{\rm min}(t_\star)\right) \,.
    \label{eq:numericalDelta}
\end{equation}
Here we introduced $t_{\rm min}(f)$, defined as the time at which $f$ equals the minimal frequency defined in Eq. \eqref{eq:fmin}: $f=f_{\rm min}(t_{\rm min})$. Note that a $t_{\rm min}(f)$ that is larger than $t_\star$ corresponds to frequencies that are too low to be emitted by the network before its decay. We therefore expect a cut-off to the spectrum at the frequency $f=f_{\rm min}(t_\star)$. The resulting SGWB is plotted in Fig.~\ref{fig:numericalNus} for $t_\star\ge t_0$, which corresponds to domain walls that survive until the present time. Therein one may see that this spectrum peaks at a frequency of $\alpha f \approx 10^{-18} \, \rm Hz$, corresponding to that of the gravitational radiation emitted by domain walls collapsing at late times. A lower value of $\alpha$ will shift the spectrum towards higher frequencies, without affecting its shape. The spectrum falls off steeply for frequencies larger than the peak frequency, until it smoothly transitions to a flatter --- but still rapidly decreasing --- part that corresponds to the GWs emitted in the radiation era. We will discuss the case of a $\delta$-shaped PDF in more detail in section \ref{sec:ResAnalytic}.

If we now consider the more general probability density function in Eq.~\eqref{eq:PDFnu}, the GWs are emitted with frequencies $f \geq f_{\rm min}$. This means that, for any frequency $f$, the SGWB will only have contributions from the domains that have decayed after $t_{\rm min}(f)$ . We may then write the spectrum $\Omega_{\rm gw}(f)$ as:
\begin{eqnarray}
    \Omega_{\rm gw}(f) & = & -(\nu+1)\frac{8\pi G}{3H_0^2} \mathcal{F}\tilde{c}\sigma\left(\frac{\alpha f}{2}\right)^{\nu+1}  \label{eq:omegaNuNumerical} \\ 
    & \times & \int_{t_{\rm min}}^{t_{\rm f}} \left( \frac{a(t)}{a_0}\right)^{3-\nu}\bar{v}(t)L(t)^{\nu-1} dt \nonumber \, .  
\end{eqnarray}
Fig.~\ref{fig:numericalNus} shows  the SGWB generated by domain walls with $t_\star \ge t_0$ in a radiation + $\rm \Lambda$CDM background for this more general PDF and for different values of the spectral index $\nu$. The key feature of this spectrum is also a peak at the low-frequency end, corresponding to walls decaying late into the matter-era --- its position  is also roughly given by $\alpha f \approx 10^{-18}\, \rm Hz$. However, an increase in $\nu$ represents a power shift towards higher frequencies and is thus associated with a decrease of the peak amplitude. In fact, the amplitude vanishes asymptotically in the $\nu \rightarrow -1$ limit. For smaller values of $\nu$, on the other hand, the spectrum becomes steeper and asymptotically approaches that produced by the Dirac-$\delta$ PDF in Eq.~(\ref{eq:numericalDelta}) in the $\nu\to-\infty$ limit, as this corresponds to concentrating all the energy produced by the network at any given time around $f_{\rm min}$. Moreover, as before, a smaller value of $\alpha$ will shift the spectrum to higher frequencies, without affecting its shape. Naturally, as will be shown in more detail later in this section, if the domain walls are biased and decay before the present time, the spectrum would have a cut-off at a frequency $f_{\rm min} (t_\star)$ and, therefore, the peak would appear at higher frequencies.

\subsection{Analytical approximation to the domain wall SGWB}\label{sec:ResAnalytic}
Assuming a power law evolution for the scale factor and that the domain wall network is in a linear scaling regime --- which, as we discussed, is a good approximation deep into the radiation and matter eras--- it is possible to obtain an analytical approximation for the SGWB generated by domain wall networks that survive until a time $t_\star$. As we will show, this analytical approximation not only provides an excellent fit for the spectrum obtained numerically --- thus enabling a simpler and faster computation --- but also aids in understanding better its shape.

The following spectrum is obtained if $p(f)$ is given by a $\delta$-function as in Eq. \eqref{eq:pdfDelta}:

\begin{eqnarray}
    \Omega_{\rm gw}(f) & = &\mathcal{A}_\lambda \mathcal{F} \mathcal{C}_\lambda^4 \frac{8\pi G \sigma}{3H_0^2 (1-\lambda)} \nonumber \\ 
    & \times & \left( \frac{\alpha \xi_\lambda f}{2\mathcal{C}_\lambda}  \right)^{\frac{4\lambda-1}{\lambda-1}} \theta \left( f - f_{\rm min}(t_\star)\right)\,,
    \label{eq:analyticalDelta}
\end{eqnarray}
where we have $\mathcal{A}_{\rm r}=0.102$ and $\xi_{\rm r}=1.53$ during the radiation era and $\mathcal{A}_{\rm m}=0.064$, $\xi_{\rm m}=1.62$ for the matter era.
The spectrum given in Eq. \eqref{eq:analyticalDelta} is shaped as a power-law, scaling proportionally to $f^{-5}$ in the matter era and $f^{-2}$ in radiation domination. As the GWs emitted during the matter era correspond to the low frequency part of the spectrum, while higher frequencies are emitted in the radiation era, this result is consistent with the general shape of the spectrum obtained numerically (as plotted in Fig. \ref{fig:numericalNus}). We will return to this result later in this section.

\subsubsection{Walls decaying in the radiation era}
Let us now consider the case of the PDF in Eq. \eqref{eq:PDFnu} and consider  domain walls that disappear, due to the effect of bias, at some time $t_\star<t_{\rm eq}$. In this case, assuming scaling, it is possible to obtain an explicit expression for $t_{\rm min}(f)$ --- which yields $t_{\rm min}=[2\mathcal{C}_{\rm r}/(\alpha \xi_{\rm r} f) ]^{2}$ --- and to use it to perform the integration in Eq.~(\ref{eq:defOmega}) analytically. We find

\begin{eqnarray}
    \Omega^{\rm r}_{\rm gw}(f,a_\star,\nu) & = & \frac{32\pi}{3} \frac{\nu+1}{\nu+3}\mathcal{F} G \sigma \frac{\Omega_{\rm r}^{1/2}}{H_0}\mathcal{A}_{\rm r}\label{eq:SpecRad}\\ & \hspace{-2.4cm} \times & \hspace{-1.4cm} \left(\frac{\alpha\xi_{\rm r} f}{4H_0\Omega_{\rm r}^{1/2}}\right)^{-2} \left[ 1-\left(\frac{\alpha\xi_{\rm r} f}{4H_0\Omega_{\rm r}^{1/2}}\frac{a_\star}{a_0}\right)^{\nu+3}\right]\nonumber\,, 
\end{eqnarray}
which is valid for $\nu \neq \nu_{\rm pole}^{\rm r}=-3$. Note that, if $\nu > \nu_{\rm pole}^{\rm r}$, the radiation era spectrum will be dominated by the second term in Eq.~(\ref{eq:SpecRad}) and will therefore scale proportionally to $f^{\nu+1}$. This means that, in this limit, the SGWB is dominated by the last GWs emitted by the domain wall network over all the frequency range. On the other hand, for all values of $\nu$ smaller than $\nu^{\rm r}_{\rm pole}$, the first term in Eq.~(\ref{eq:SpecRad}) dominates and the spectral shape will be independent of $\nu$, showing a power-law decay with $\Omega_{\rm gw} \propto f^{-2}$ similar to that obtained for the $\delta$-function PDF. 

For $\nu=-3$, on the other hand, we find that:
\begin{eqnarray}
    \Omega_{\rm gw}^{\rm r}(f,a_\star, \nu_{\rm pole}^{\rm r}) & = &  \frac{64\pi}{3}\mathcal{F} G \sigma \frac{\Omega_{\rm r}^{1/2}}{H_0}\mathcal{A}_{\rm r} \label{eq:SpecRadLog} \\ & \times & \left(\frac{\alpha \xi_{\rm r} f}{4H_0\Omega_{\rm r}^{1/2}} \right)^{-2} \log \left[ \frac{\alpha \xi_{\rm r} f  }{4H_0\Omega_{\rm r}^{1/2}}\frac{a_\star}{a_0}\right] \nonumber \, .
\end{eqnarray}
In this case, the spectrum will again scale asymptotically as $f^{-2}$ for sufficiently high frequencies, although in this case the logarithmic term in Eq. (\ref{eq:SpecRadLog}) results in a slightly shallower decay in the low-frequency part.

For a biased domain wall network that disappears in the radiation era, the SGWB spectrum peaks at a frequency

\be
\alpha f^{\rm r}_{\rm peak}=\begin{cases}
    \frac{4H_0\Omega_{\rm r}^{1/2}}{\xi_{\rm r}} \left[-\frac{2}{\nu+1}\right]^{1/(\nu+3)} \left(\frac{a_0}{a_\star}\right), & \text{for }\nu\neq -3\,, \\
    \frac{4H_0\Omega_{\rm r}^{1/2}}{\xi_{\rm r}} \mathrm{e}^{1/2} \frac{a_0}{a_\star}, & \text{for } \nu = -3\,,
  \end{cases}
\ee 
which corresponds to a peak amplitude of

\be 
    \Omega^{\rm r}_{\rm peak} = \frac{32\pi}{3 } \mathcal{F} G \sigma  \frac{\Omega_{\rm r}^{1/2}}{H_0}\mathcal{A}_{\rm r}  \left(\frac{a_\star}{a_0}\right)^2\left[-\frac{2}{\nu+1}\right]^{-2/(\nu+3)} \label{eq:SpecRadpeak}\,,
\ee
for $\nu\neq\nu_{\rm pole}^{\rm r}$ and

\begin{equation}
    \Omega^{\rm r}_{\rm peak}=\frac{32\pi}{3 \mathrm{e}} G\sigma \mathcal{F} \frac{\Omega_{\rm r}^{1/2}}{H_0} \mathcal{A}_{\rm r} \left(\frac{a_\star}{a_0}\right)^2 \,,
\end{equation}
for $\nu=-3$. It is interesting to note that the amplitude of the peak, except for a factor $2[-2/(\nu+1)]^{-2/(\nu+1)}$ for $\nu\neq -3$ and $2/\mathrm{e}$ for $\nu=-3$, coincides with the total energy density in GWs in units of the critical density of the universe in Eq.~(\ref{eq:totrhorad}). These factors, except in the $\nu \to -1$ limit (in which significant power is shifted towards high frequencies) are, roughly, of order unity, which shows that, in general, the SGWB is dominated by the emission at late times.

\subsubsection{Walls decaying in the matter era}

For domain wall networks that survive through the radiation-matter transition and disappear at a time $t_\star\ge t_{\rm eq}$, we will assume, for simplicity, that the transition between these two eras happens suddenly at $t_{\rm eq}$. We assume then the universe to be effectively radiation (matter) dominated, with $\lambda=1/2$ ($2/3$) for $t< t_{\rm eq}$ ($t\ge t_{\rm eq}$) and that the network is in a scaling regime characterized by the parameters $\xi_{\rm r}$ ($\xi_{\rm m}$) and $\bar{v}_{\rm r}$ ($\bar{v}_{\rm m}$). During the matter era, $t_{\rm min}=[2\mathcal{C}_{\rm m}/(\alpha\xi_{\rm m} f)]^3$ and so the SGWB is given by\footnote{Technically, to compute the matter era contribution, the integration in Eq.~(\ref{eq:defOmega}) should start at $t_{\rm i}=\rm{max}(t_{\rm min},t_{\rm eq})$. Here, to maintain the analytical expressions presented as simple as possible, we always set $t_{\rm i}=t_{\rm min}(f)$. We found that, if $\nu<-6$, the matter-era contribution changes its slope at $f_{\rm min} (t_{\rm eq})$ and becomes steeper for $f>f_{\rm min} (t_{\rm eq})$, if one takes $t_{\rm i}=\rm{max}(t_{\rm min},t_{\rm eq})$. However, since in this case the SGWB spectrum is dominated by the walls that decay in the radiation era in this frequency range, this has, in general, no significant impact on the final results. Note however that taking $t_{\rm i}=\rm{max}(t_{\rm min},t_{\rm eq})$ is necessary if one is interested in computing solely the contribution of the walls that decay in the matter era.}
\begin{eqnarray}
    \Omega^{\rm m}_{\rm gw}(f,a_\star,\nu) & = & \Omega^{\rm r}_{\rm gw}(f,a_{\rm eq},\nu)+12\pi \frac{\nu+1}{\nu+6}\mathcal{F}G\sigma\label{eq:SpecMat} \\ & \hspace{-4.1cm}\times & \hspace{-2.3cm} \frac{\Omega_{\rm m}^{1/2}}{H_0}\mathcal{A}_{\rm m} \left( \frac{\alpha \xi_{\rm m} f}{3H_0\Omega_{\rm m}^{1/2}}\right)^{-5}\left\{1- \left[\frac{\alpha\xi_{\rm m}f}{3H_0\Omega_{\rm m}^{1/2}}\left(\frac{a_\star}{a_0}\right)^{1/2}\right]^{\nu+6}\right\} \nonumber\,,
\end{eqnarray}
which has a pole for $\nu \equiv \nu_{\rm pole}^{\rm m}=-6$. As was the case for the contribution of the radiation-era walls, if $\nu>\nu_{\rm pole}^{\rm m}$, this spectrum will scale proportionally to $f^{\nu+1}$, while for $\nu<\nu_{\rm pole}^{\rm m}$ we have that $\Omega_{\rm gw}^{\rm m}\propto f^{\nu_{\rm pole}^{\rm m}+1}=f^{-5}$ as in the case of the Dirac-$\delta$ PDF. For $\nu=-6$, we have an additional logarithmic dependency on the frequency just as in the radiation-dominated case:

\begin{eqnarray}
    \Omega^{\rm m}_{\rm gw}(f) & = & \Omega^{\rm r}_{\rm gw}(f,a_{\rm eq},\nu)+60 \pi \mathcal{F} G\sigma  \frac{\Omega_{\rm m}^{1/2}}{H_0}\mathcal{A}_{\rm m} \label{eq:logMatterSpec} \\ & \times & \left(\frac{\alpha \xi_{\rm m} f}{3H_0\Omega_{\rm m}^{1/2}}\right)^{-5} \log \left[ \frac{ \alpha \xi_{\rm m} f} {3H_0\Omega_{\rm m}^{1/2}}\left(\frac{a_\star}{a_0}\right)^{1/2}\right] \nonumber\,.
\end{eqnarray}

In this case, the peak is located at a frequency
\be
\alpha f^{\rm m}_{\rm peak}=\begin{cases}
    \frac{3H_0\Omega_{\rm m}^{1/2}}{\xi_{\rm m}} \left[-\frac{5}{\nu+1}\right]^{1/(\nu+6)} \left(\frac{a_0}{a_\star}\right)^{1/2}, & \text{for }\nu\neq -6\,, \\
    \frac{3H_0\Omega_{\rm m}^{1/2}}{\xi_{\rm m}} \mathrm{e}^{1/5} \left(\frac{a_0}{a_\star}\right)^{1/2}, & \text{for } \nu = -6\,,
  \end{cases}
\ee 
which corresponds to a peak amplitude of

\be 
    \Omega^{\rm m}_{\rm peak} = 12\pi\mathcal{F} G \sigma  \frac{\Omega_{\rm m}^{1/2}}{H_0}\mathcal{A}_{\rm m}  \left(\frac{a_\star}{a_0}\right)^{5/2}\left[-\frac{5}{\nu+1}\right]^{-5/(\nu+6)}\,,
\ee
for $\nu\neq\nu_{\rm pole}^{\rm m}$ and

\begin{equation}
    \Omega^{\rm m}_{\rm peak}=\frac{12}{\mathrm{e}}\pi \mathcal{F} G\sigma \frac{\Omega_{\rm m}^{1/2}}{H_0}\mathcal{A}_{\rm m} \left(\frac{a_\star}{a_0}\right)^{5/2} \,,
\end{equation}
for $\nu=-6$. As before, these amplitudes follow closely the total amplitude in GWs in units of the critical density in Eq.~(\ref{eq:torrhomat}), except for factors of $5[-5/(\nu+1)]^{-5/(\nu+6)}$ for $\nu\neq -6$ and $5/e$ for $\nu=-6$.

\begin{figure} 
	         \begin{minipage}{1.\linewidth}  
                \rotatebox{0}{\includegraphics[width=1\linewidth]{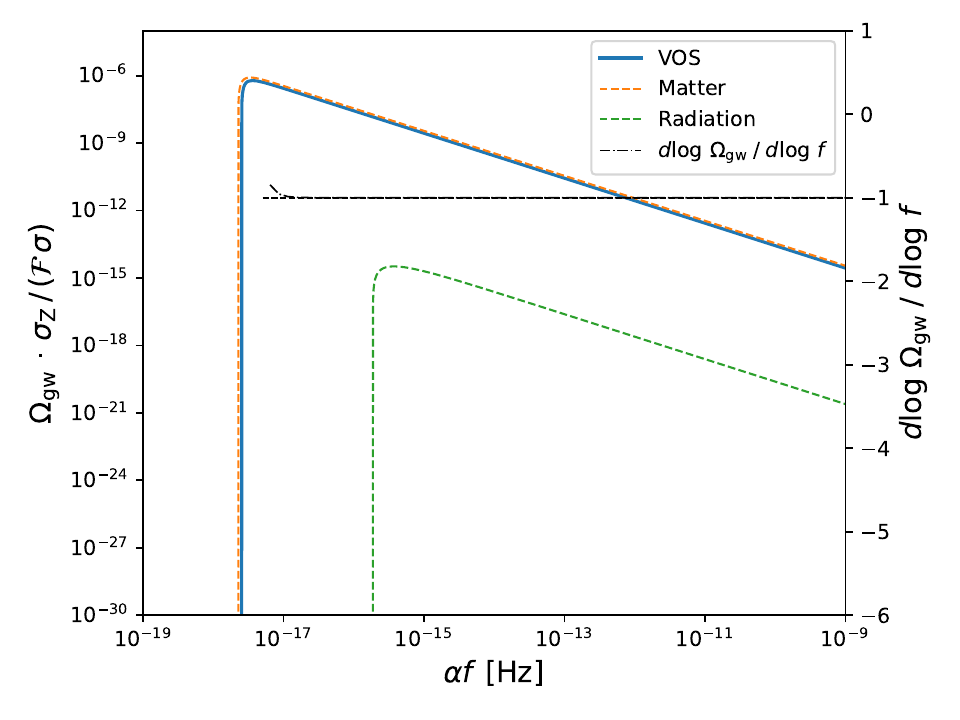}}
               \rotatebox{0}{\includegraphics[width=1\linewidth]{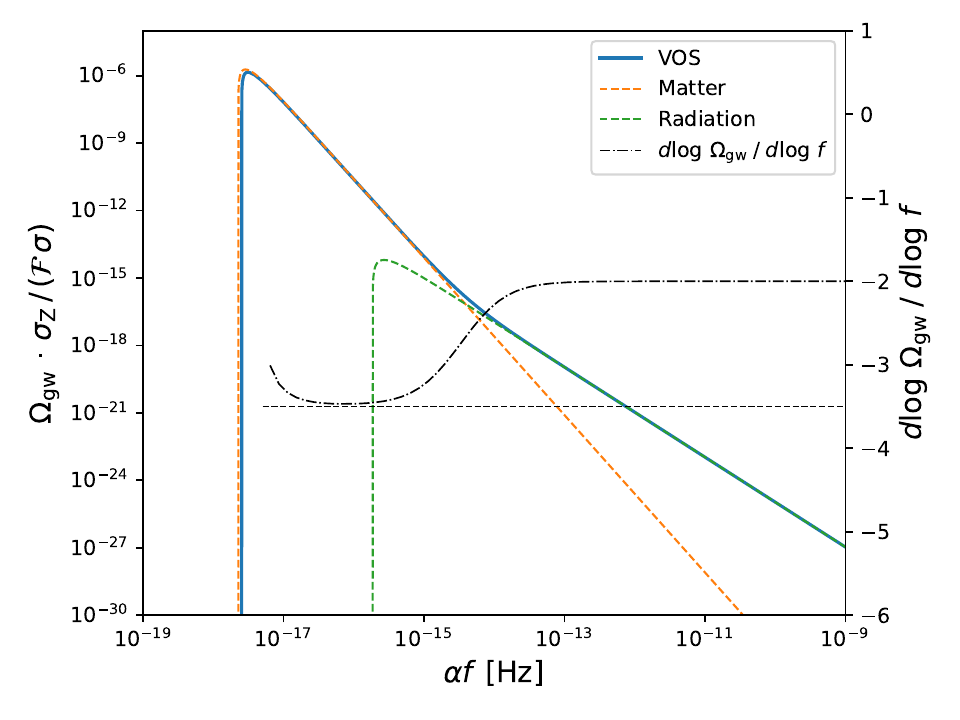}}
               \rotatebox{0}{\includegraphics[width=1\linewidth]{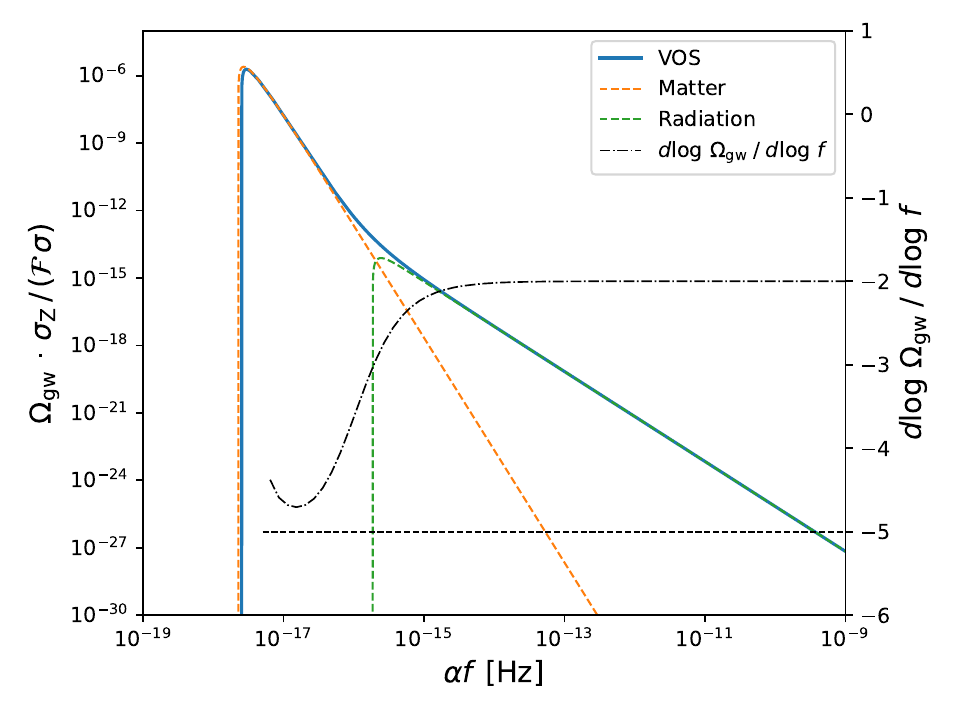}}
         \end{minipage}
	\caption{Analytical approximation to the SGWB generated by domain wall networks that decay, due to the effect of bias, at $t_\star\ge t_0$. These plots display the full spectra obtained numerically (solid lines) as well as the analytical approximations for the contributions generated during the matter (dashed orange lines) and radiation era (dashed green lines) for  $\nu=-2$ (top panel), $\nu=-4.5$ (middle panel) and $\nu = -7$ (bottom panel). Each panel additionally displays the spectral index $d\Omega/d\log f$ of the numerical spectrum (dash-dotted lines) and that predicted by the analytical approximation during the matter era (black dashed line).}
	\label{fig:analytical}
\end{figure}

We may then divide the results produced by our model into three different regimes:
\begin{enumerate}
\item For $-3 \leq \nu < -1$, the spectra generated in the radiation- and matter-dominated eras have equal slopes. As a consequence, the spectrum will be globally dominated by walls that have decayed more recently, while the contribution from walls decaying early in the evolution of the network is negligible. This behaviour is clearly displayed in the top panel of Fig.~\ref{fig:analytical}, where we plot the analytical approximation to the the SGWB generated by domain walls derived here, for $\nu=-2$ and $t_\star\geq t_0$, alongside the full spectrum obtained numerically. This regime, which corresponds to having significant power emitted in higher frequencies, particularly includes the above mentioned case in which $\nu \rightarrow -1$, corresponding to a flat spectrum. However, as may be seen in the Eqs. (\ref{eq:SpecRad},~\ref{eq:SpecMat}), the $\nu+1$ pre-factor will lead to a suppression of the peak amplitude of the spectrum in this limit. 
\item For $-6 < \nu < -3$, the radiation era spectrum scales as $f^{-2}$, while the matter-era contribution becomes steeper with decreasing $\nu$. An example of the SGWB generated by walls in this regime, with $\nu=-4.5$ and $t_\star\geq t_0$, is depicted in the middle panel of Fig. \ref{fig:analytical}. Therein, one may see clearly that the contributions of the radiation and matter eras become clearly distinguishable in this regime and that, as the frequency increases, the spectrum is dominated by the GW emissions of domains emitted at progressively earlier times.
\item For all $\nu \leq -6$, the spectrum will scale proportionality to $f^{-5}$ in matter- and $f^{-2}$ in radiation-domination, as shown in the bottom panel of Fig. \ref{fig:analytical}, where we plot the SGWB generated by domain walls with $\nu=-7$ and $t_\star\geq t_0$. In this regime, the spectrum follows very closely that obtained for a $\delta$-shaped PDF [Eq. \eqref{eq:analyticalDelta}], since this corresponds to concentrating most of the GW power emitted around $f_{\rm min}$. The latter can therefore be regarded as an asymptotic limit for low values of $\nu$, in agreement with the numerical results presented in Fig. \ref{fig:numericalNus}.
\end{enumerate}

The plots in Fig.~\ref{fig:analytical} show that the analytical approximations derived here provide an excellent fit to the SGWB obtained numerically over the whole range of the spectrum and for all values of $\nu$. This figure additionally shows the spectral index of the full numerically-obtained spectrum, defined as $d\log \Omega_{\rm gw}/d\log f$. One may see that, for the cases with a distinct scaling of the spectrum with frequency in the radiation- and matter-dominated eras, the transition between the two characteristic slopes is smooth. This is a consequence of the fact that the radiation-matter transition is a slow continuous process rather than an instantaneous one (as we have assumed for the purpose of deriving the analytical approximations). In fact, for low values of $\nu$, the slope predicted by the analytical approximation in matter-domination is never fully reached and the spectrum will be shallower than predicted in the low-frequency regime. However, even in this case, our analytical approximation still provides a remarkably good fit to the numerical results.

Finally, it is worth noting that, with on average only about one wall existing per Hubble volume in a scaling network, the number of collapse events in the recent cosmic history should be relatively low. This raises questions regarding the stochastic nature of the gravitational wave background generated by domain walls decaying in the very recent past and this may result, in fact, in a ``popcorn'' signal or in individual resolvable bursts of GWs.

\subsection{The contribution of bias}\label{sec:resBias}

When bias becomes relevant to domain wall dynamics, it renders the whole network unstable and all the walls evaporate within about a Hubble time. In this process, domain walls emit further gravitational radiation and this emission should also contribute to the SGWB. We did not take this final emission into consideration in the previous sections. Here, we briefly discuss this additional contribution to the SGWB.

As was the case for the GWs emitted throughout the evolution of the network, in this case, it is not known \textit{a priori} how much of the energy of the network is converted into gravitational radiation and how this energy is distributed in frequency in this almost instantaneous decay process. As a matter of fact, bias may have an impact on the emission of scalar radiation and the distribution of GWs in frequency may in fact be altered since, in this case, there is also a decay of super-horizon domains. Here, for simplicity and given all the unknowns, we will assume that both the PDF and $\mathcal{F}$ will be the same as in the case of the domains that collapse while the network is stable. As the final decay of a biased network occurs in roughly one Hubble time, we will assume it may be treated as instantaneous on cosmic time scales. In this approximation, all the energy stored in the network is assumed to be released at the time of decay, $t_\star$. Then, the contribution of this decay to the GW spectrum as measured today will be given by the Eq.~(\ref{eq:gw-density}) and~(\ref{eq:spectral-gw-density}), but with the energy loss term replaced by~\cite{Dunsky:2021tih}:
\begin{equation}
    \left.\frac{d\rho}{dt}\right|_{\rm loss, bias}=\rho \delta(t-t_\star)\,.
\end{equation}
The integration of a term of this form will yield an additional contribution of the form $\Omega_{\rm gw}\propto f^{\nu+1}$. For a network decaying in the radiation-dominated era, with $t_\star < t_{\rm eq}$, the full spectrum may be obtained by multiplying the second term in Eq.~(\ref{eq:SpecRad}) by the following term:
\be 
\left[ 1+ \frac{\nu+3}{2 \mathcal{A}_{\rm r} \xi_{\rm r}} \theta\left(f-f_{\rm min}(a_\star)\right)\ \right]\,.
\ee
As to the $\nu=-3$ case, one has to add the following term to Eq. \eqref{eq:SpecRadLog}
\be
\frac{32}{3}\pi \mathcal{F} G\sigma \frac{\Omega_{\rm r}^{1/2}}{H_0 \xi_{\rm r}} \left(\frac{\alpha\xi_{\rm r} f}{4H_0 \Omega_{\rm r}^{1/2}}\right)^{-2}\, ,
\ee
in order to account for the contribution of bias.
If on the other hand $t_\star\ge t_{\rm eq}$, the last term in Eq.~(\ref{eq:SpecMat}) should be corrected by a factor of
\begin{eqnarray}
     \left[ 1 + \frac{\nu+6}{3\mathcal{A}_{\rm m} \xi_{\rm m}} \theta \left(f-f_{\rm min}(a_\star)\right)  \right]\,,
\end{eqnarray}
for $\nu\neq-6$ and the term
\be
20\pi \mathcal{F}G\sigma \frac{\Omega_{\rm m}^{1/2}}{H_0 \xi_{\rm m}}\left(\frac{\alpha\xi_{\rm m} f}{3H_0 \Omega_{\rm m}^{1/2}}\right)^{-5}
\ee
should be added to Eq. \eqref{eq:logMatterSpec} for $\nu=-6$.

\begin{figure} 
	         \begin{minipage}{1.\linewidth}  
               \rotatebox{0}{\includegraphics[width=1\linewidth]{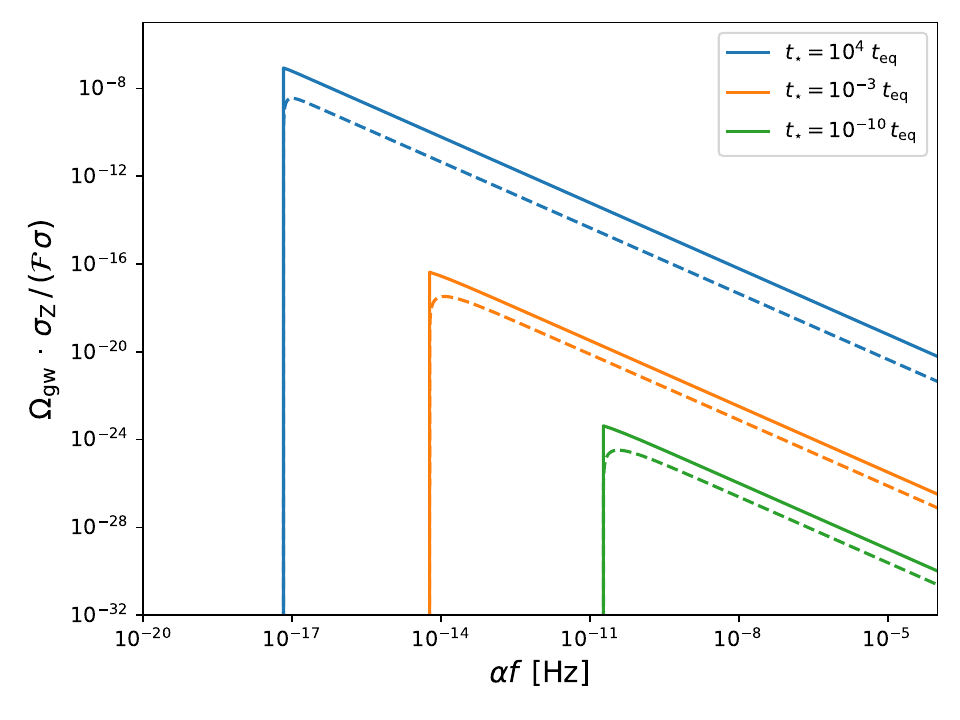}}
               \rotatebox{0}{\includegraphics[width=1\linewidth]{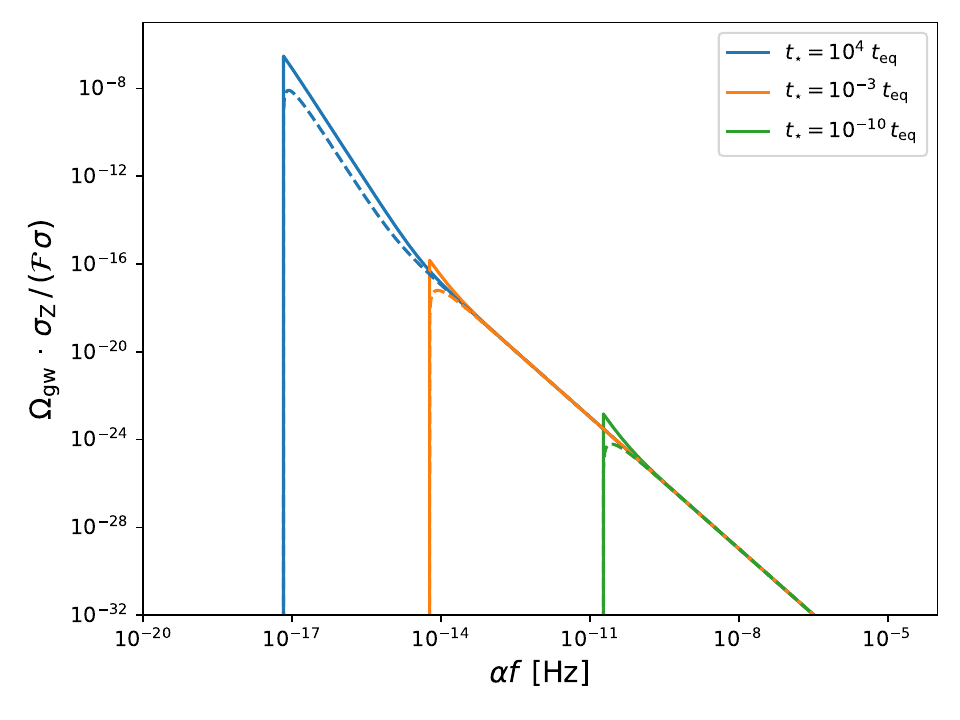}}
               \rotatebox{0}{\includegraphics[width=1\linewidth]{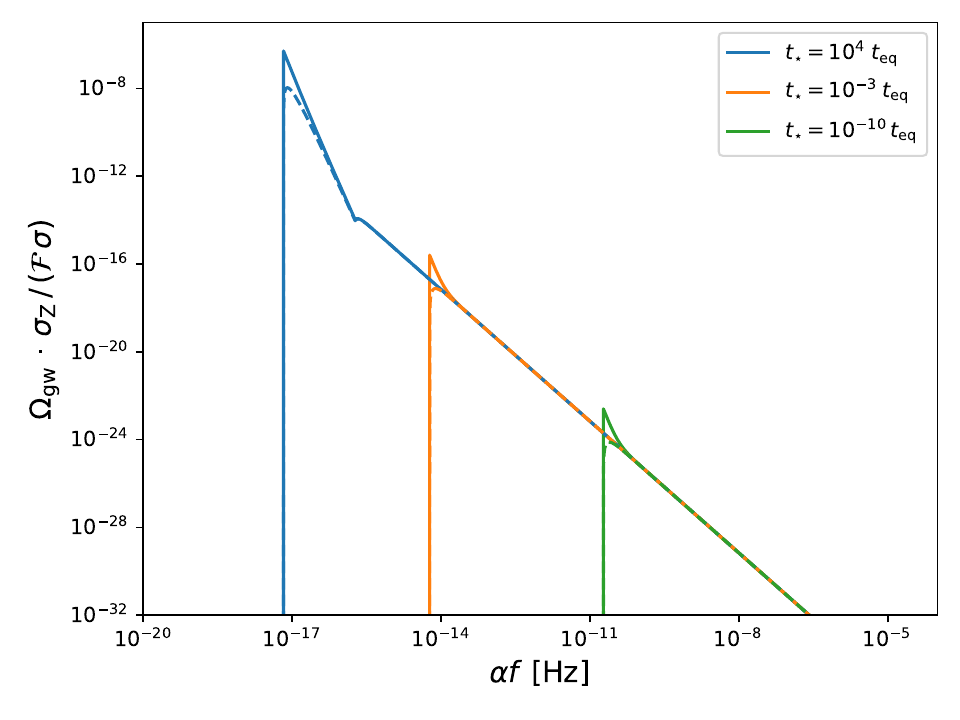}}
         \end{minipage}
	\caption{Stochastic gravitational wave background generated by biased domain wall networks. Each panel displays the analytical approximation the total SGWB spectrum generated by a biased network for different decay times $t_\star$ including the contribution of the final decay of the network (solid lines), alongside the spectrum generated by the network while it is stable up until time of decay (dashed lines). The top, middle and bottom panels correspond, respectively, to values of $\nu=-2$, $\nu = -4.5$ and $\nu = -7$.}
	\label{fig:decaySpec}
\end{figure}

Fig.~\ref{fig:decaySpec} displays examples of the full spectrum expected for a biased network decaying at different instants of time $t_\star$ (given as a fraction of the time of matter-radiation equality $t_{\rm eq}$) and for different values of $\nu$.
These plots show that the sudden decay of the network has the potential to increase the peak amplitude of the SGWB by up to one order of magnitude. This is to be expected, as the energy density in the network at any time is larger --- but roughly of the same order --- than the energy density lost by the network in a Hubble time.  Moreover, as we have seen, the contribution of the decay of the biased network (in our simplified model) always scales as $f^{\nu+1}$. As a consequence, for $\nu \geq \nu_{\rm pole}^{\rm r}$ (the case in which the ``scaling'' spectrum also scales as $f^{\nu+1}$), the spectrum is globally dominated by the gravitational waves emitted during the decay of the network. This case is displayed in the top panel of Fig.~\ref{fig:decaySpec}. Besides the peak getting sharper --- a consequence of assuming that the decay would happen instantaneously --- this case is not distinguishable from the spectrum generated by a stable domain wall network with a larger tension $\sigma$ or an increased $\mathcal{F}$ (as these would lead to an increase of the amplitude). On the other hand, for $\nu < \nu_{\rm pole}^{\rm r}$, the part of the spectrum emitted before bias comes into effect scales as $f^{-2}$ for GWs sourced in the radiation era, as explained in section \ref{sec:ResAnalytic}. This means that, for $t_\star<t_{\rm eq}$, the contribution of the decay of the network caused by bias dominates the spectrum only in a small frequency band close to the peak, while towards higher frequencies it is dominated by the contribution emitted while the network is stable. A similar situation arises for $t_\star\ge t_{\rm eq}$ when $\nu < \nu_{\rm pole}^{\rm m}$. These situations, as can be seen in the middle and bottom panels of Fig. \ref{fig:decaySpec}, lead to a distinct sharper peak with a different slope that becomes more pronounced the smaller $\nu$ gets, which may be regarded as a potential observational signature of bias.

As to the total energy density of gravitational radiation measured by an observer at a time $t_{\rm f}$, to account for the final collapse of the network, we need to include an additional contribution of $\rho$ at $t=t_\star$ and to take into account that there is no emission of GWs for $t>t_\star$. We then have:

\begin{eqnarray}
    \rho_{\rm gw}(t_{\rm f}) =  \sigma\frac{\mathcal{F}}{t_\star}\frac{\mathcal{A}_\lambda}{4\lambda-1}\left[\left(1+\frac{4\lambda-1}{\mathcal{A}_\lambda\xi_\lambda}\right)\left(\frac{t_\star}{t_{\rm f}}\right)^{4\lambda}\right.\\
    \left.  -  \frac{t_\star}{t_{\rm i}}\left(\frac{t_{\rm i}}{t_{\rm f}}\right)^{4\lambda}\right]\,,\nonumber
\end{eqnarray}
for $t_{\rm f}\ge t_\star$, while Eq.~(\ref{eq:totalrhogw}) remains valid if $t_{\rm f}<t_\star$. Assuming that $t_{\rm f} \gtrsim t_\star \gg t_{\rm i}$, the total GW energy density is enhanced by a factor of $\sim 7$, if bias comes into effect during the radiation era, and a factor of $\sim 17$ for $t_\star \ge t_{\rm eq}$. For times $t_{\rm f}>t_\star$, however, there is no new gravitational wave emission and, therefore, the energy density of gravitational radiation is merely redshifted due to expansion. Note that the enhancement factors predicted here should be taken as an upper limit since a more realistic non-instantaneous description of the decay of the wall network would necessarily lead to smaller values and to spectra with rounder peaks and a smaller maximum amplitude. Also, we have ignored the potential impact of bias on $\mathcal{F}$ and the PDF as this is currently unknown, but if these are significantly impacted this contribution would be significantly affected too.

\section{Comparison with previous computations}\label{sec:comparisonLiterature}

The shape of the SGWB generated by domain wall networks computed here is different from other predictions in the literature~\cite{Hiramatsu:2013qaa,Kitajima:2015nla,Krajewski:2016vbr,Chen:2020wvu,Dunsky:2021tih,Ferreira:2022zzo,Badger:2022nwo}. These predictions all stem from the results of Ref.~\cite{Hiramatsu:2013qaa}, wherein the spectrum of GWs is measured in field theory simulations of domain wall networks (see also~\cite{Gleiser:1998na,Hiramatsu:2010yz,Kawasaki:2011vv,Hiramatsu:2012sc} for earlier estimates and~\cite{Saikawa:2017hiv} for a review). In~\cite{Hiramatsu:2013qaa}, the authors propose a broken power-law shape for the spectrum, with the peak located at a frequency $f_{\rm peak}$ corresponding to the horizon scale at the time of disappearance of the network, if walls are biased, or the present time for standard walls. For $f>f_{\rm peak}$, simulations indicate that $\Omega_{\rm gw} \propto f^{-1}$ in the radiation era --- which would correspond to a PDF for GW emission with a spectral index of $\nu=-2$ in our case --- however the authors state that the exact dependence of $\Omega_{\rm gw}$ on $f$ is not straightforward to estimate. For $f<f_{\rm peak}$, the dependence of $\Omega_{\rm gw}$ on $f$ was not accurately determined in the simulations and the authors claim that causality requires that $\Omega_{\rm gw} \propto f^{3}$, as was shown to be the case for the SGWB generated by bubble collisions in first order phase transitions~\cite{Caprini:2009fx} (see however \cite{Cutting:2018tjt,Cai:2019cdl}). Here, we obtained a much steeper spectrum for $f<f_{\rm peak}$ --- as is, in fact, also found for the SGWB generated by cosmic strings~\cite{sousaStochasticGravitationalWave2013,sousaStochasticGravitationalWave2014,Sousa:2020sxs}, especially for fast-decaying small cosmic string loops~\cite{sousaStochasticGravitationalWave2014} --- and this steepness is a consequence of assuming that one may associate a well defined characteristic frequency to the collapsing domain walls. Note however that our predictions for this spectrum (and the predictions for that generated by cosmic string loops in the literature) do not apply to superhorizon scales: we always consider a minimum frequency of emission that corresponds to a sub-horizon mode as, in fact, the GWs are sourced, in this case, by domains that are sub-horizon at the time of collapse. In reality, a less steep cut-off would be expected, and thus this framework cannot be expected to provide an accurate description of the spectra for frequencies smaller than that of the peak. As a matter of fact, a smooth transition towards a $f^3$ spectrum might be expected to occur in this regime, but the precise nature of this transition is a topic that warrants further investigation.

Our predictions for the shape of the spectrum qualitatively agree with the results in the literature for frequencies higher than that of the peak: in both cases the spectrum is dominated by the last GWs emitted and decreases as a power law. The biggest disagreement between our predictions and those of~\cite{Hiramatsu:2013qaa} actually appears in the normalization of the spectrum. By comparing our results in Eq.~(\ref{eq:SpecRadpeak}) to those therein, we find that the latter include an additional factor of $\sim G \sigma /H(t_\star)$. This factor may be quite significant and may have implications for the observational constraints derived using the results of~\cite{Hiramatsu:2013qaa}. In~\cite{Hiramatsu:2013qaa}, the normalization of the spectrum is determined by assuming that the energy density in gravitational waves remains constant over time, as predicted in~\cite{Hiramatsu:2010yz}. As we have argued in Sec.~\ref{sec:resTotalEnergy}, this should not be the case: in an expanding background, if $\mathcal F$ is a constant, the total gravitational radiation energy density measured by an observer at a time $t$, after an initial period of quick growth, necessarily decreases as $t$ increases. 
A constant GW energy density would necessarily require that the fraction of the energy lost by the network into GWs would increase over time (with $\mathcal{F} \propto t$, if linear scaling of $L$ with cosmic time is assumed) and, as previously explained, such scenario would necessarily have to be transient since the condition $\mathcal{F}\le 1$ must be satisfied at all times\footnote{Notice that, given that the domain walls will also lose energy through other channels,  $\mathcal{F} = 1$ will never be attained in practice.}. 
The measurements in Ref.~\cite{Hiramatsu:2013qaa} seem to be consistent with a constant $\rho_{\rm gw}$, and these are admittedly affected by several systematic uncertainties. In fact, these simulations are not able to resolve the dynamics of the domain walls in the  ultrarelativistic regime, in which the dominant contribution of domain walls to the SGWB is expected to be generated, nor do they account for the GW backreaction that would necessarily affect domain wall dynamics, especially for values of $\mathcal{F}$ not much smaller than unity --- a necessary condition for domain walls to  provide a significant contribution to the SGWB background.

Our results about the evolution of $\rho_{\rm gw}$ are based on minimal assumptions: they are independent of both the characteristic scale and the spectrum of emission of GWs. The only underlying assumption is that the energy that goes into GWs is a constant fraction of that lost by the domain wall network. Although this assumption may not hold throughout the whole evolution of the network, it should be an excellent approximation during the final stages of the evolution of the domain walls, when most of the contribution to the SGWB is generated. The assumptions in~\cite{Hiramatsu:2010yz,Hiramatsu:2013qaa} are different, as the authors assume that the GWs are sourced by domain walls that are only mildly relativistic and that did not fully detach from the Hubble flow. Therein, following~\cite{Vilenkin:1984ib}, the authors use the quadrupole formula to estimate the power emitted by a domain wall with a typical size of $R$:

\be 
P \sim G\sigma^2 R^8 f^6\,.
\ee 
Then, assuming that the frequency is determined by $R$, so that $f\sim R^{-1}$, one has

\be 
P \sim G\sigma^2 R^2\,.
\ee 
The energy in a  volume $L^3$, containing a domain wall with area $L^2$, is then estimated in~\cite{Hiramatsu:2010yz,Hiramatsu:2013qaa} to be $E\sim P \cdot t \sim P\cdot L \sim G\sigma ^2 L^3$, so that $\rho_{\rm gw}\sim E/L^3$ remains constant over time. This is the result that the authors use as a basis to derive the normalization of the spectra. In the last step of this derivation, the radius of the domain wall $R$ was identified with the characteristic length $L$ of the network and linear scaling of the network was assumed. However, \textit{one cannot have a cake and eat it}: emission of radiation by a domain wall, in any form, would necessarily lead to a decrease of their size and, hence, the identification of $R$ and $L$ is questionable in this case. More formally, one would write the power emitted per unit volume by the domain wall network in the form of gravitational waves as

\be 
\frac{d\rho_{\rm gw [e]}}{dt}=P n\,,\quad \mbox{where} \quad n=\rho/E_{\rm w}
\ee
is the wall number density, and $E_{\rm w}$ and $P$ are, respectively, the average energy of a single wall and the average power emitted per wall in the form of GWs. Estimating $P$ using the quadrupole formula, one obtains 

\be 
P \sim G\sigma^2 L^2\,,
\ee 
which then implies 

\be 
\frac{d\rho_{\rm gw [e]}}{dt}\sim\frac{G\sigma^2}{L} = \frac{\rho}{\tau}\,,
\label{eq:}
\ee
where $\tau = (G \sigma)^{-1}$ defines a characteristic time for the conversion of a constant fraction of the energy of the network into GWs (which for reasonable values of domain wall tension is always much larger than the age of the universe). Notice that, in this case, we are not following the evolution of a single wall and, therefore, no identification between $R$ and $L$ is necessary. By integrating and maintaining the assumption that the network is in linear scaling, one would in fact obtain a constant energy density. Note however that there is an underlying assumption when one proceeds this way: walls are assumed to only decay at this rather slow rate and to be extremely long-lived. This can only apply to walls that are part of the network and not to collapsing domain walls. The quadrupole formula, in fact, cannot capture what happens in the last ultra-relativistic stages of collapse and the potential bursts triggered by domain wall self-interactions which can lead to a smaller time-dependent characteristic time $\tau \propto L$ for the production of GWs  (which for a scaling network would imply that $\tau \propto H^{-1}$). In this case, we would then have that the averaged power emitted per unit volume during the collapse scales proportionally to $L^{-2}$, precisely as predicted for the energy loss rate of the domain wall network (cf. Eq.~(\ref{eq:VOSLoss})). The radiation coming from the slowly moving walls that are part of the network should be subdominant when compared to that emitted in the last stages of the evolution of the many collapsing domains that exist at any time. The fact that the energy lost by a network of domain walls during its evolution is so well described by the rapid collapse of the domains that detach from the Hubble flow after entering the horizon~\cite{avelinoParameterfreeVelocitydependentOnescale2020} provides strong evidence for this.

\section{Conclusions}\label{sec:conclusions}

In this paper we investigated the power spectrum of the SGWB produced by domain wall networks. We assumed that the main source of energy loss is the collapse, on a timescale of roughly one Hubble time, of domains walls that detach from the Hubble flow and that these walls disappear quickly after, leaving behind a contribution to the gravitational wave and scalar radiation backgrounds. These basic assumptions were shown to be sufficient to estimate the energy density associated with the SGWB generated by domain walls, as a function of two crucial parameters of our model: the efficiency of GW emission and the domain wall tension. We modelled the scale dependence of the average distribution in frequency of the GWs emitted by domain walls in the final collapse stages using a simple model characterized by two further parameters: a minimum frequency inversely proportional to the characteristic length of the network at the time of collapse and a constant power spectrum slope. We also investigated the limiting case where all the emission happens at the minimum frequency. For all these cases we were able to compute the power spectrum of the SGWB generated by domain walls as observed at the present time. We also developed an analytical approximation assuming perfect linear scaling in both radiation and matter eras. We have shown that the power spectrum of the SGWB background generated by domain walls is a monotonic decreasing function of the frequency whose peak and slope depends significantly on the two parameters characterising the average frequency distribution of the GWs generated by domain walls at the time of emission. This should constitute a strong motivation for an in-depth characterisation of the GWs generated by collapsing domain walls. Probing the dynamics of domain walls in the ultra-relativistic regime --- during which their main contribution to the SGWB is expected to be generated --- will probably require a new generation of field theory simulations, capable of dealing with the major challenge posed by the enormous Lorentz contraction of the width of ultra-relativistic domain walls. We have also investigated the case of biased domain walls, considering both the contribution to the SGWB generated while the network is stable and the contribution associated to the sudden decay of the domain walls.

\begin{acknowledgments}
L. S. thanks Ricardo Z. Ferreira, Oriol Pujolas and Marek Lewicki for enlightening discussions about this subject.  D. G. is supported by FCT -- Funda\c{c}\~{a}o para a Ci\^{e}ncia e a Tecnologia through the PhD fellowship 2020.07632.BD. L. S. is supported by FCT through contract No. DL 57/2016/CP1364/CT0001. Funding for this work has also been provided by FCT through the research grants UIDB/04434/2020 and UIDP/04434/2020 and through the R \& D project 2022.03495.PTDC -- \textit{Uncovering the nature of cosmic strings}.

\end{acknowledgments}
 
\bibliography{walls}
 	
\end{document}